\documentclass[journal,final]{IEEEtran}
\usepackage{amsmath}
\usepackage{amsthm}
\usepackage{amssymb}
\usepackage{amsfonts}
\usepackage{soul,color}
\usepackage{url}
\usepackage{pgfplots}
\pgfplotsset{compat=1.18}
\usepackage{tikz}
\usepackage{bm,cite}
\usetikzlibrary{arrows.meta, positioning}
\usetikzlibrary{decorations.pathreplacing} 
\usepackage[acronym,shortcuts]{glossaries}
\newacronym{rff}{RFF}{radio frequency fingerprint}
\newacronym{mcsi}{M-CSI}{Micro-CSI}
\newacronym{ofdm}{OFDM}{orthogonal frequency division multiplexing}
\newacronym{dft}{DFT}{discrete Fourier transform}
\newacronym{cir}{CIR}{channel impulse response}
\newacronym{rf}{RF}{radio frequency}
\newacronym{csi}{CSI}{channel state information}
\newacronym{glrt}{GLRT}{generalized likelihood ratio test}
\newacronym{iot}{IoT}{internet of things}
\newacronym{auc}{AUC}{area under the curve}
\newacronym{det}{DET}{detection error tradeoff}
\newacronym{wls}{WLS}{weighted least squares}
\newacronym{rffi}{RFFI}{radio frequency fingerprint identification}
\newacronym{lt}{LT}{likelihood test}
\newacronym{awgn}{AWGN}{additive white Gaussian noise}
\newacronym{cdf}{CDF}{cumulative distribution function}

\title{From Identification to Authentication for Micro-CSI RF Fingerprinting in OFDM Systems \thanks{Corresponding author: Matteo Varotto.\\
Matteo Varotto is with the Department of Information Engineering, University of Padova, 35131 Padua, Italy (e-mail:
matteo.varotto.3@phd.unipd.it).
\\
Stefano Tomasin is with the Department of Information Engineering and the Department of Mathematics of 
University of Padova, 35131 Padua, Italy, and also with National Interuniversity Consortium for Telecommunications (CNIT), 43124 Parma, Italy
(e-mail: stefano.tomasin@unipd.it).
\\
The authors gratefully acknowledge Dr. He Chen from the Chinese University of Hong Kong for providing the conceptual basis and early insights that inspired this research.

}

}
\author{
Matteo Varotto and Stefano Tomasin
}
\begin{document}
\maketitle 
\begin{abstract}
We consider a scenario where a legitimate user (Alice) authenticates itself to an authenticator (Bob) by transmitting \ac{ofdm} pilots, from which the authenticator extracts the \ac{mcsi} fingerprint and compares it against a stored reference via a \ac{lt}-based test. We introduce a new spoofing attack, where two adversarial devices collude to first jointly estimate the \ac{mcsi} fingerprints of Alice and Bob and then construct a forged signal able to break the authentication mechanism with high probability, limited only by noise effects on the estimates. We derive approximate closed-form distributions of the authentication test statistic under both the legitimate and spoofing hypotheses, enabling the derivation of false alarm and misdetection probabilities in closed-form. We then validate our analytical results against \ac{mcsi} fingerprints extracted from experimental data. The results reveal that, given sufficient pilot observations or an equivalent noise statistic between Bob and the attackers, the latter can always drive the test statistics to a random classifier, vanishing the security of \ac{mcsi}-based authentication.

\glsresetall
\end{abstract}
\section{Introducion}
The proliferation of wireless devices, especially in the \ac{iot} world, has highlighted the need for lightweight, physical-layer security systems to authenticate devices without cryptographic systems. \Ac{rff} has emerged as a promising solution, using the hardware impairments generated during the manufacturing process as a unique feature for identification \cite{mattia_survey, survey2,survey3}. Unlike cryptographic mechanisms adopted at higher layers, \ac{rff} requires no key management overhead and works transparently at the physical layer, making it very attractive, especially for \ac{iot} resource-constrained environments. In this paper, we focus on micro-signals embedded in \ac{csi} measurements, which serve as stable, device-specific \ac{rf} fingerprints in \ac{ofdm} systems \cite{chen,chen2}. These micro-signals are called \ac{mcsi} and arise from \ac{rf} circuitry imperfections at each device, mainly from the power amplifier embedded in each \ac{rf} chain. Crucially, because they originate from hardware, they require special techniques to mimic or suppress, representing a good system for identification.

Most of the existing literature on \ac{rff} such as \cite{mattia_survey, survey2,survey3,rffi_identification} has been used for the {\em identification task}: given a received signal, determine which device among a known set transmitted it. In this case, it is assumed that devices are acting honestly and do not aim to alter identification to impersonate (or spoof) other devices. Part of the literature \cite{chen, wang, smartphone,novel_authentication,rffi_identification,csi_paper} instead has proposed \ac{rffi} also as an {\em authentication mechanism}, i.e., to identify devices even in a scenario where an attacking device aims to impersonate a legitimate device. In such a context, a recurring assumption is that hardware fingerprints are inherently difficult to replicate or forge, making them intrinsically secure identifiers. This assumption is often invoked to argue that \ac{rff}-based systems provide a secure authentication mechanism \cite{privacy_fingerprint,adc_authentication}. However, in a security context, the adversary can deliberately attempt to evade the defense's authentication system by doing smarter attacks than simply transmitting with its own equipment. 
As noted in \cite{smartphone}, wireless transmitter fingerprinting systems are vulnerable to forgery attacks carried out with software-defined radio equipment. 
Another relevant example of smart attacks is presented in \cite{rffi_ai_spoofing}, where a generative adversarial neural network generates synthetically adversarial samples that fool the \ac{rff}-based identification.  However, this claim is rarely subjected to rigorous adversarial analysis \cite{rffi_adversarial} or is based on the assumption of naive attackers \cite{rffi_zero_trust, rffi_auth_AI,chen}.

In this paper, we provide a rigorous analytical study of the security limits of \ac{mcsi}-based authentication in \ac{ofdm} systems. We consider that a legitimate user Alice is authenticated by a verifier Bob by transmitting \ac{ofdm} pilots, from which Bob extracts the \ac{mcsi} fingerprint and compares it against a stored reference via a \ac{lt}-based test statistic. We consider a smart attack implemented by two colluding devices, named Trudy and Chunk. They eavesdrop the signal received during the pilot exchanges between Alice and Bob to jointly estimate the \ac{mcsi} fingerprints of Trudy and Alice. Then, Trudy constructs a forged signal that pre-cancels both her own hardware signature and her channel response toward Bob, so that the fingerprint extracted by Bob resembles Alice's stored reference. We note that the colluder Chuck is needed to acquire the \ac{mcsi} of Trudy, which is then pre-compensated during the attack. We analytically characterize the performance of this system by deriving approximate distributions of the authentication test statistic under both the legitimate and the spoofing hypotheses, enabling a closed-form description of the fundamental security limits of \ac{mcsi}-based authentication. We then validate our analytical results against \ac{mcsi} fingerprints extracted from experimental data. The results reveal that, given sufficient pilot observations or an equivalent noise statistic between Bob and the attackers, the latter can always drive the test statistics to a random classifier, vanishing the security of \ac{mcsi}-based authentication. 

The rest of the paper is organized as follows. Section~\ref{sec::system} describes the system model. Section~\ref{sec::authentication} describes the authentication process performed by Bob. Section~\ref{sec::attacker} describes the strategy of the attackers. Section~\ref{sec::test_function} derives a closed-form expression for the security parameters of the system. Section~\ref{sec::numerical} displays the numerical results, while Section~\ref{sec::conclusion} concludes the paper.

\section{System Model}
\label{sec::system}
\begin{figure}
    \centering
    \includegraphics[width=1\linewidth]{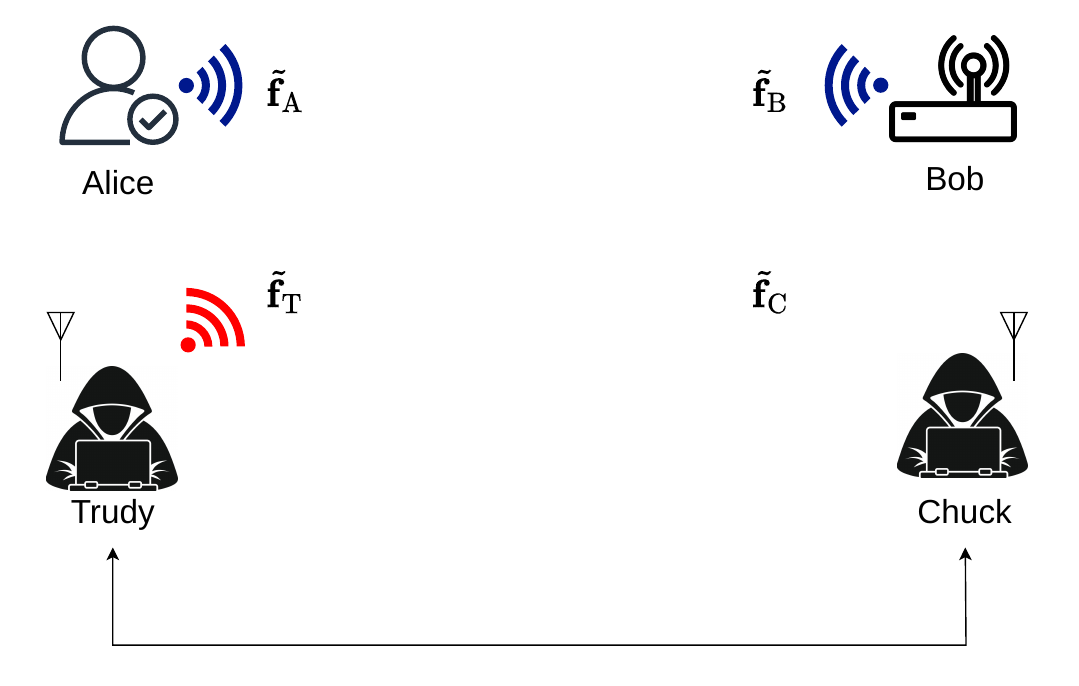}
    \caption{Schematic representation of the considered scenario.}
    \label{fig::scenario}
\end{figure}

We consider the wireless network of Fig.~\ref{fig::scenario}, where a device Bob is receiving messages that are transmitted by either Alice (the legitimate user) or Trudy (the attacker). Bob aims at determining the source of the received message based on the properties of the received physical-layer signal. In particular, we want to exploit the \ac{mcsi} features of the transmitting device for authentication.
To this end (and also to support communication targets) Alice and Trudy transmit an \ac{ofdm} signal that includes unit-power pilot signals on some subcarriers, from which Bob can estimate the channel (and the \ac{mcsi}). The pilot structure is assumed to be public, thus also known to the intruder. In its intrusion, Trudy is supported by a colluding device, denoted as Chuck. Moreover, we assume that Bob also transmits the pilot signals, enabling Trudy and Chuck to estimate the \ac{mcsi} of this device.

\subsection{Channel and Signal Model}
In an \ac{ofdm} system with $K$ subcarriers, when a device transmits a pilot symbol $x_k \in \mathbb{C}$ with power $P_B$ on subcarrier $k=1,\dots, K$, the received signal on subcarrier $k$ is
\begin{equation}
       r_k = c_k \, x_k + \omega_k, \quad
    \omega_k \sim \mathcal{CN}(0, \sigma^2_\omega),
    \label{eq:received_signal}
\end{equation}
where ${c}_k \in \mathbb{C}$ is the \ac{csi} on subcarrier $k$ and $\omega_k$ is \ac{awgn}. Dividing $r_k$ by the known pilot symbol $x_k$, the \ac{csi} estimate is obtained as ${c}_k = r_k / x_k$.

Being any physical device affected by manufacturing imperfections, we model its received signal as shaped by both the transmitter's and receiver's imperfections. Let ${f}^{TX}_k \in 
\mathbb{C}$ and ${f}^{RX}_k \in \mathbb{C}$ are the micro-CSI individual fingerprints 
induced by the transmitter's and receiver's RF hardware imperfections respectively on subcarrier $k$. The frequency-domain \ac{csi} estimated by a receiver is modeled as \cite{chen}
\begin{equation}
    {c}_k = {h}_k(1 + {f}^{TX}_k)(1 + {f}^{RX}_k), \quad k=1, \ldots, K
    \label{eq:signal_model}
\end{equation}
where ${h}_k \in \mathbb{C}$ is the component due to transmission on air. 
Since the \ac{mcsi} fingerprints are significantly smaller than the ideal signal 
($|{f}^{TX}_k|, |{f}^{RX}_k| \ll 1$), the second-order product 
 ${f}^{TX}_k {f}^{RX}_k$ is negligible, and we can approximate \eqref{eq:signal_model} as:
\begin{equation}
    {c}_k \approx {h}_k(1 + {f}^{TX}_k + {f}^{RX}_k), \quad k=1, \ldots, K.
    \label{eq:signal_model_approx}
\end{equation}
We then define the composite fingerprint ${f}_k \triangleq 1+ {f}^{TX}_k + {f}^{RX}_k$.

Stacking channel parameters of all $K$ subcarriers into column vectors of $K$ complex numbers
 ${\bm{c}}$, ${\bm{h}}$, ${\bm{f}}^{TX}$, and ${\bm{f}}^{RX}$,
where the $k$-th entry of each vector corresponds to subcarrier $k$,  and defining ${\bm{f}} = {\bm{f}}^{TX} + {\bm{f}}^{RX}$, 
\eqref{eq:signal_model_approx} can be rewritten as 
\begin{equation}
    {\bm{c}} \approx {\bm{h}} \circ (\bm{1} + \bm{f}),
    \label{eq:signal_model_vector}
\end{equation}
where $\bm{1} \in \mathbb{R}^{K \times 1}$ is the all-ones column vector 
and $\circ$ denotes element-wise multiplication.

\subsection{Signal Space Projection and Channel Estimation}

The key observation underlying \ac{mcsi} estimation is that the physical channel $\bm{h}$ and the composite fingerprint $\bm{f}$ occupy different regions of the time domain. Let the $N$-size \ac{cir} column vector $\bm{g} \in \mathbb{C}^N$, with entry $n$ representing the channel tap at delay $nT_s$, where $T_s$ is the sampling period. In \ac{ofdm} systems, synchronization algorithms align the received frame with the strongest tap, which in line-of-sight scenarios corresponds to the central tap of the \ac{cir}. Taps arriving before this central tap are circularly shifted to the end of the delay window, and therefore appear at negative indices mod $N$. 

As a result, in a typical indoor environment, $\bm{g}$ has significant energy concentrated in only the $N_p \ll N$ taps spanning symmetrically around this central tap, reflecting the physical multipath delays of the propagation environment. By contrast, the hardware-induced fingerprint perturbation does not share this sparse structure: in the frequency domain it arises as an element-wise product, which corresponds to a convolution in the time domain, dispersing its energy across many taps well beyond $N_p$.

This separation motivates a signal-space projection approach, introduced in \cite{chen} and summarized here. 
We define the subspace $\mathcal{V} \subset \mathbb{C}^K$ spanned by the \ac{dft} columns corresponding to delays $nT_s$ for $n \in \mathcal{L} = \{-N_p, \ldots, N_p\}$ mod $N$, with associated projection matrix
\begin{equation}
    \bm{A} = \bm{F}_{\mathcal{L}}(\bm{F}_{\mathcal{L}}^*
    \bm{F}_{\mathcal{L}})^{-1}\bm{F}_{\mathcal{L}}^* \in \mathbb{C}^{K \times K},
    \label{eq:projection_matrix}
\end{equation}
where $\bm{F}_{\mathcal{L}} \in \mathbb{C}^{K \times |\mathcal{L}|}$ collects the \ac{dft} columns indexed by $\mathcal{L}$.

Projecting \ac{csi} ${\bm{c}}$ onto $\mathcal{V}$ retains the channel contribution ${\bm{h}}$, which lies entirely within $\mathcal{V}$ by assumption, while suppressing the fingerprint perturbation, which lies predominantly outside $\mathcal{V}$. The resulting smoothed channel estimate is:
\begin{equation}
    \hat{{\bm{h}}} = \bm{A}{\bm{c}}+\bm{\hat{z}} \quad \bm{\hat{z}} \sim \mathcal{CN}\left(0, \sigma^2_{z}\right),
    \label{eq:channel_estimate}
\end{equation}
where $\bm{\hat{z}}$ is the channel estimation error.

Once $\hat{{\bm{h}}}$ is available, the \ac{mcsi} fingerprint is estimated by element-wise division of the estimated \ac{csi},  removing the channel contribution and leaving the hardware-induced perturbation as the quantity of interest, as detailed in the following subsection.

\subsection{Composite Fingerprint Estimation}

Given $\hat{{\bm{h}}}$, we can estimate the composite fingerprint over each subcarrier $k$ as follows :
\begin{equation}
    \hat{{f}}_k = \frac{{c}_k}{\hat{{h}}_k} =
    \frac{{h}_k(1 + {f}^{TX}_k + {f}^{RX}_k) + {z}_k}
    {\hat{{h}}_k},
    \label{eq:extraction_derivation}
\end{equation}
where the approximation is obtained by ignoring the estimation error on $\hat{\bm h}$.
Generalizing this to the entire bandwidth, the composite fingerprint $\hat{{\bm{f}}}$ is estimated by element-wise division of the \ac{csi} vector by the smoothed channel estimate $\hat{{\bm{h}}}$:
\begin{equation}
    \hat{{\bm{f}}} = {\bm{c}} \oslash
    \hat{{\bm{h}}}.
    \label{eq:fingerprint_extraction}
\end{equation}
Note that the extracted fingerprint is invariant to 
the wireless channel ${\bm{h}}$ regardless of the transmitter's position 
or the propagation environment. 
\paragraph*{Estimation Approximation}
Given the assumption $\hat{\bm{h}}\approx \bm{h}$ note that the estimation process \eqref{eq:fingerprint_extraction} gives
\begin{equation}
\hat{\bm{f}}\approx \bm{1} + {\bm{f}}^{TX} 
    + {\bm{f}}^{RX} +\boldsymbol{\varepsilon},
    \label{eq::approximation}
\end{equation}
where $\boldsymbol{\varepsilon}$ is the residual \ac{awgn} after composite fingerprint estimation. Note that being in a TX-RX chain the composite fingerprint is composed by the two individual fingerprints ${\bm{f}}^{TX}$ and ${\bm{f}}^{RX}$.

\subsection{Attacker Model}

We consider two attackers, Trudy and Chunck, that collude  to impersonate Alice and bypass Bob's authentication test. Both Trudy and Chuck can receive signals. They passively observe the pilot signals transmitted by Alice and Bob. Only Trudy is an active transmitter, as it is the device that injects the spoofing signal toward Bob. Trudy and Chuck share all collected measurements via an error-free link. We also assume that both attackers have full knowledge of the system parameters, including the pilot sequences, the number of subcarriers $K$, the composite fingerprint estimation process \eqref{eq:fingerprint_extraction}, and their own noise statistics. The details of how Trudy and Chuck jointly exploit these capabilities to construct the spoofing signal are described in Section~\ref{sec::attacker}.

Alice is unaware of the existence of Trudy, and in particular Alice does not know the fingerprint of either Trudy or Chuck. We also assume Bob transmits pilot symbols, enabling Trudy and Chuck to estimate their individual fingerprint.

\section{Physical Layer Authentication With \ac{mcsi}}
\label{sec::authentication}
The authentication framework involves two distinct phases. The enrollment phase is executed only once per legitimate device. In this phase, Bob collects multiple pilot transmissions from Alice and stores a reference composite fingerprint estimate $\bm f_{\text{Alice}}$ that will be used for future authentication. The authentication phase is then repeated at every received message, where Bob extracts the composite fingerprint from the incoming transmission and compares it against the stored reference using a suitable test.

\subsection{Enrollment Phase}
To store a reference/ground truth of Alice's \ac{mcsi}, Bob collects $N_{\text{E}}$ \ac{ofdm} symbols from Alice over the static channel $\bm{h}_{AB} = [h_{AB, 1}, \ldots, h_{AB, K}]^T$. On subcarrier $k$ and observation $i=1,\dots,N_E$, Bob receives after pilot division:
\begin{multline}
c^{(i)}_{B,k} = h_{AB,k}\left(1 + f_{A,k} + f_{B,k}\right) +\, z^{(i)}_{B,k},\\
z^{(i)}_{B,k} \sim \mathcal{CN}(0, \sigma^2_B).
\label{eq:csi_enrollment}
\end{multline}
By using \eqref{eq::approximation} we model the stored reference as
\begin{multline}
    f_{\text{Alice},k} \approx  1+{f}_{\text{A},k} + {f}_{\text{B},k} +\varepsilon_{E,k}, \\
    \varepsilon_{E,k} \sim \mathcal{CN}\left(0, \frac{\sigma^2_B}{N_{\text{E}} |h_{AB,k}|^2}\right).
    \label{eq:f_alice}
\end{multline}
A detailed derivation of \eqref{eq:f_alice} is presented in Appendix~\ref{app::1}. Note that we assume that Trudy does not transmit in this phase and Alice does not know its fingerprint.

\paragraph*{Remark} We note that the pilots transmitted by Alice should be known to Bob to enable channel estimation. Furthermore, we assume here that they are public, i.e., also known to Trudy and Chuck. If this is not the case, there is a pre-shared secret between Alice and Bob, on which then the authentication mechanism is based. Here instead we consider a scenario where authentication is obtained without pre-shared secrets.

\subsection{Authentication Phase}
At each authentication round, Bob receives $N_A$ \ac{ofdm} symbols and must decide whether they originate from Alice (hypothesis $\mathcal{H}_0$) or from an attacker (hypothesis $\mathcal{H}_1$). The decision is taken on the basis of the estimated fingerprint from the received signal, that we indicate with $\hat{f}_k$.

When Alice transmits a pilot symbol $x_k$, the \ac{csi} observed by Bob on subcarrier $k$ at observation $i = 1, \ldots, N_A$ is (from \eqref{eq:signal_model_approx})
\begin{multline}
    {c}^{\mathcal{H}_0,(i)}_{k} = {h}_{AB,k}(1 + {f}_{A,k}+ {f}_{B,k}) + z^{(i)}_{B,k}, \\
    z^{(i)}_{B,k} \sim \mathcal{CN}(0, \sigma^2_B),
    \label{eq::csi_alice}
\end{multline}
where ${h}_{AB,k}$ is the channel frequency response between Alice and Bob on subcarrier $k$. By applying the estimation  \eqref{eq:fingerprint_extraction} to each observation and averaging over the $N_A$ received \ac{ofdm} symbols, the composite fingerprint estimate is:
\begin{multline}
    \hat{f}^{\mathcal{H}_0}_{k} \approx  1 + f_{A,k} + f_{B,k} + \varepsilon_{A,k}, \\
    \varepsilon_{A,k} \sim \mathcal{CN}\left(0, \frac{\sigma^2_B}{N_A |h_{AB,k}|^2}\right).
    \label{eq:f_auth}
\end{multline}

When Trudy transmits a spoofing signal $s_{k} \in \mathbb{C}$, (whose design is detailed in Section~\ref{sec::attacker}), the \ac{csi} observed by Bob at \ac{ofdm} symbol $i$ becomes:
\begin{equation}
    {c}^{\mathcal{H}_1,(i)}_{k} = {h}_{TB,k}(1 + {f}_{T,k}+ {f}_{B,k})  s_{k} + z^{(i)}_{B,k},
    \label{eq:csi_bob_raw}
\end{equation}
where ${h}_{TB,k}$ is the channel frequency response between Trudy and Bob and ${f}_{T,k}$ is Trudy's hardware fingerprint. In both cases, Bob compares the extracted composite fingerprint against the stored reference $\bm{f}_{\text{Alice}}$ via the following test statistic.

 Following the \ac{lt} framework presented in \cite{baracca} and defining the total noise variance per subcarrier as $\nu^2_k$, we define the decision statistic  as the noise-weighted squared Euclidean distance across all $K$ subcarriers, i.e.,
\begin{equation}
    \Psi (\hat{f}) = 2\sum_{k=1}^{K} \frac{|\hat{f}_k - f_{\text{Alice},k}|^2}{\nu^2_k}.
    \label{eq:glrt}
\end{equation}
Then, the authentication is a decision process with the following test:
\begin{equation}\label{eq:classifier}
{\hat{\mathcal H}(\bm{{f}})} = \begin{cases}
\mathcal H_0 &\mbox{if } \Psi(\bm{{f}}) \leq \tau, \\
\mathcal H_1 &\mbox{if } \Psi(\bm{{f}}) > \tau,\\
\end{cases}
\end{equation}
where $\tau$ is the suitable threshold.

The performance of the test are measured by the false alarm and misdetection probabilities, defined as 
\begin{equation}
P_{\rm FA} = {\mathbb P}[{\hat{\mathcal H}(\bm{{f}})} = \mathcal H_1 |\mathcal H(\bm{{f}}) = \mathcal H_0]\,,
\end{equation}
\begin{equation}
P_{\rm MD} = {\mathbb P}[{\hat{\mathcal H}(\bm{{f}})} = \mathcal H_0 |\mathcal H(\bm{{f}}) = \mathcal H_1],
\end{equation}
where $\mathcal H(\bm{{f}})$ is the true condition (legitimate or spoofed) of the received \ac{mcsi} $\bm{{f}}$.
The choice of $\tau$ provides a fundamental trade-off between $P_{\rm MD}$ and $P_{\rm FA}$. A higher $\tau$ reduces $P_{\rm FA}$ while increasing $P_{\rm MD}$. In practice, being  $P_{\rm FA}$ related only to samples belonging to $\mathcal H_0$, $\tau$ is typically set to satisfy a specific $P_{\rm FA}$ constraint.
\paragraph*{Common Receiver Derivation}
Let $\mathcal{X}$ be a generic transmitter with hardware fingerprint $f_{X,k}$. The CSI observed by Bob on subcarrier $k$ is:
\begin{equation}
    c_k = h_{XB,k}(1 + f_{X,k} + f_{B,k}) + z_k.
    \label{eq:common_csi}
\end{equation}
Applying the estimation \eqref{eq:fingerprint_extraction}, the composite fingerprint estimate is: 
\begin{equation}
    {f}_k \approx 1 + f_{X,k} + f_{B,k} + \varepsilon_k,
    \label{eq:common_fp}
\end{equation}
where $\varepsilon_k$ is residual noise after channel cancellation. The stored reference from enrollment is \eqref{eq:f_alice}. The residual entering the test statistic \eqref{eq:glrt} for a generic transmitter $\mathcal{X}$ is therefore:
\begin{equation}
    {f}_k - f_{\text{Alice},k} = (f_{\mathcal X,k} - f_{A,k}) + (\varepsilon_k - \varepsilon_{\text{Bob},k}).
    \label{eq:common_residual}
\end{equation}
Since Bob is fixed across both enrollment and authentication, $f_{B,k}$ appears identically in both terms and cancels exactly, regardless of who is transmitting or how the fingerprint $f_{X,k}$ is structured.

\section{Colluding Attack of The  \\
M-CSI PLA Mechanism}
\label{sec::attacker}

We design a spoofing signal $\bm{s} \in \mathbb{C}^K$ such that
the fingerprint extracted by Bob, ${f}$, is recognized as legitimate.

Given the estimate $\hat{h}_{TB,k}$, the estimate ${f}_{T,k}$ of its own 
fingerprint, and the estimate ${f}_{\text{A},k}$ of Alice's fingerprint computed before, Trudy constructs the spoofing 
signal on each subcarrier $k$ as:
\begin{equation}
s_{k} = \frac{{1+{f}}_{A,k}}{{(1+{f}}_{T,k})  \hat{h}_{TB,k}}.
    \label{eq:spoofing_signal}
\end{equation}
We note that when the estimates are perfect, inserting \eqref{eq:spoofing_signal} into \eqref{eq:csi_bob_raw} we obtain \eqref{eq::csi_alice}, i.e., the \ac{csi} estimated by Bob is the same as when Alice is transmitting. Therefore with this attack we have $P_{\rm MD}=1- P_{\rm FA}$.
In fact, when these two conditions are met, we have:
\begin{align}
    c^{\mathcal{H}_1}_{k} &= h_{TB,k}(1 + f_{T,k} + f_{B,k})  \frac{1+f_{A,k}}{(1+f_{T,k})  h_{TB,k}} + z_{B,k} \notag \\
    &= \frac{1 + f_{T,k} + f_{B,k}}{1+f_{T,k}}(1+f_{A,k}) + z_{B,k} \notag \\
    &= \left(1 + \frac{f_{B,k}}{1+f_{T,k}}\right)(1+f_{A,k}) + z_{B,k} \notag \\
    &\approx (1+f_{B,k})(1+f_{A,k}) + z_{B,k} \notag \\
    &\approx h_{AB,k}(1 + f_{A,k} + f_{B,k}) + z_{B,k} = c^{\mathcal{H}_0}_{k}.
    \label{eq:perfect_spoof}
\end{align}

However, Trudy faces several challenges in constructing $\bm{s}$. First, Trudy's own \ac{rf} hardware produces a fingerprint ${\bm{f}}_T \in \mathbb{C}^K$ into its transmission,
which Bob will extract instead of $f_{\text{Alice}}$ unless Trudy pre-cancels it. To do this, we will show that Trudy needs Chuck, whose role is to send collected composite fingerprint estimations to Trudy so it is possible to correctly estimate the individual fingerprint of the devices in the security scenario.
Second, Trudy must cancel the effect of her channel $\bm{h}_{TB}$ on her transmitted signal, which requires estimating $\bm{h}_{TB}$ from Bob's pilot transmissions. 
\subsection{Trudy-Bob Channel Estimation}

Trudy passively observes $L_T$ pilot transmissions from Bob, receiving on
subcarrier $k$ and observation $l$ the signal, which includes
Bob's hardware fingerprint $f_{B,k}$ and Trudy's own fingerprint $f_{T,k}$:
\begin{equation}
    r^{(l)}_{T,k} = h_{TB,k}(1 + f_{B,k} + f_{T,k}) x^{(l)}_k + \omega^{(l)}_{T,k},
    \qquad l = 1, \ldots, L_T,
    \label{eq:trudy_observation}
\end{equation}
where $x^{(l)}_k$ is Bob's pilot symbol with power $P_B$ and
 $\omega^{(l)}_{T,k} \sim \mathcal{CN}(0, \sigma^2_T)$ is Trudy's receiver noise.
Since the composite fingerprint contribution $(1 + f_{B,k} + f_{T,k})$ acts
as a multiplicative factor on the channel, and $|f_{B,k}|, |f_{T,k}| \ll 1$ by the approximation of Section~\ref{sec::system}, Trudy applies the
signal-space projection $\bm{A}$ to estimate $h_{TB,k}$ directly from
\eqref{eq:trudy_observation}. As $\bm{A}$ is a linear operator, the
estimation error variance is unchanged with respect to the fingerprint-free
case, and the least-squares estimator after $L_T$ observations gives:
\begin{equation}
    \hat{h}_{TB,k} = h_{TB,k} + \varepsilon_{\text{est},k},
    \qquad
    \varepsilon_{\text{est},k} \sim \mathcal{CN}\!\left(0,\,
    \frac{\sigma^2_T}{L_T P_B}\right).
    \label{eq:trudy_channel_estimate}
\end{equation}
A detailed derivation of \eqref{eq:trudy_channel_estimate} is presented in
Appendix~\ref{app::2}.

\subsection{Micro-CSI Estimation}
\label{sec:minimal_case}

We now detail how Trudy obtains estimates of all device fingerprints with the
assistance of Chuck. Both Trudy and Chuck, knowing their internal
noise statistics, eavesdrop on the legitimate pilots exchanged by Alice and Bob.
Chuck forwards its measurements to Trudy via an error-free channel, allowing
Trudy to centrally solve the joint estimation problem. We consider a framework
where each link is observed $M_i$ times, with $M_1, M_2, M_3$, and $M_4$ denoting the number of pilot observations for the links $BT$, $BC$, $AT$, and
 $AC$ respectively.

\subsubsection*{WLS System}

The four unknown variables are the individual fingerprints of Bob, Alice, Trudy, and Chuck:
\begin{equation}
    \boldsymbol{\theta}_k = \left[{f}_{B,k},\, {f}_{A,k},\,
    {f}_{T,k},\, {f}_{C,k}\right]^T \in \mathbb{C}^4.
\end{equation}
Bob and Alice transmit pilots, while Trudy and Chuck act as receivers.
By applying the estimation  \eqref{eq::approximation} and averaging $M_i$ independent observations per link (following the same derivation as in Appendix~\ref{app::1}), Trudy obtains four composite fingerprint estimates. Specifically, each estimate on a link $X \to Y$ has the presence of residual noise term $\varepsilon_{XY,k} \sim \mathcal{CN}(0, \sigma^2_{XY,k} / M_i)$, where $\sigma^2_{XY,k}$ represents the single-observation noise variance. This yields the following four estimations:
\begin{align}
    f_{BT,k} &= 1 + {f}_{B,k} + {f}_{T,k} + \varepsilon_{BT,k}, \\
    f_{BC,k} &= 1 + {f}_{B,k} + {f}_{C,k} + \varepsilon_{BC,k}, \\
    f_{AT,k} &= 1 + {f}_{A,k} + {f}_{T,k} + \varepsilon_{AT,k}, \\
    f_{AC,k} &= 1 + {f}_{A,k} + {f}_{C,k} + \varepsilon_{AC,k}.
\end{align}
Since the additive constant $1$ is common to all four measurements, we define the shifted composite fingerprint estimates as $\bar{f}_{XY,k} = f_{XY,k} - 1$. Stacking these into the global system $\bar{\bm{f}}_k =
\bm{H}\boldsymbol{\theta}_k + \boldsymbol{\varepsilon}_k$ we obtain
\begin{equation} 
    \underbrace{\begin{bmatrix}
        \bar{f}_{BT,k} \\ \bar{f}_{BC,k} \\ \bar{f}_{AT,k} \\ \bar{f}_{AC,k}
    \end{bmatrix}}_{\bar{\bm{f}}_k}
    =
    \underbrace{\begin{bmatrix}
        1 & 0 & 1 & 0 \\
        1 & 0 & 0 & 1 \\
        0 & 1 & 1 & 0 \\
        0 & 1 & 0 & 1
    \end{bmatrix}}_{\bm{H}}
    \underbrace{\begin{bmatrix}
        {f}_{B,k} \\ {f}_{A,k} \\
        {f}_{T,k} \\ {f}_{C,k}
    \end{bmatrix}}_{\boldsymbol{\theta}_k}
    +
    \underbrace{\begin{bmatrix}
        \varepsilon_{BT,k} \\ \varepsilon_{BC,k} \\
        \varepsilon_{AT,k} \\ \varepsilon_{AC,k}
    \end{bmatrix}}_{\boldsymbol{\varepsilon}_k},
    \label{eq:minimal_system_4eq}
\end{equation}
with effective noise covariance
\begin{equation}
    {\boldsymbol{\Sigma}} = \mathrm{diag}\!\left(
    \frac{\sigma^2_{BT}}{M_1},\,
    \frac{\sigma^2_{BC}}{M_2},\,
    \frac{\sigma^2_{AT}}{M_3},\,
    \frac{\sigma^2_{AC}}{M_4}
    \right),
\end{equation}
where the division by $M_i$ reflects the variance reduction achieved by averaging $M_i$ independent observations per link. We assume that Chuck forwards all its composite fingerprint estimates to Trudy via an error-free channel, so that Trudy can assemble the full measurement vector $\bar{\bm{f}}_k$ and solve~\eqref{eq:wls_solution_4eq} centrally.

\subsubsection*{WLS Solution}

Since all noise variances in ${\boldsymbol{\Sigma}}$ are known and the
noise is complex Gaussian, the \ac{wls} estimator is the maximum likelihood
estimator and minimizes:
\begin{equation}
    \hat{\boldsymbol{\theta}}_k = \underset{\boldsymbol{\theta}
    \in \mathbb{C}^{4}}{\arg\min}\;
    (\bar{\bm{f}}_k - \bm{H}\boldsymbol{\theta})^H
    {\boldsymbol{\Sigma}}^{-1}
    (\bar{\bm{f}}_k - \bm{H}\boldsymbol{\theta}),
\end{equation}
whose closed-form solution is:
\begin{equation}
    \hat{\boldsymbol{\theta}}_k =
    \left(\bm{H}^H {\boldsymbol{\Sigma}}^{-1} \bm{H}\right)^{-1}
    \bm{H}^H {\boldsymbol{\Sigma}}^{-1} \bar{\bm{f}}_k.
    \label{eq:wls_solution_4eq}
\end{equation}
Defining ${\bm{J}} = \bm{H}^H {\boldsymbol{\Sigma}}^{-1}
\bm{H}$ and introducing the effective weights
\begin{equation}
    {\alpha} = \frac{M_1}{\sigma^2_{BT}}, \quad
    {\beta}  = \frac{M_2}{\sigma^2_{BC}}, \quad
    {\gamma} = \frac{M_3}{\sigma^2_{AT}}, \quad
    {\delta} = \frac{M_4}{\sigma^2_{AC}},
\end{equation}
we derive:
\begin{equation}
    {\bm{J}} =
    \begin{bmatrix}
        {\alpha} + {\beta} & 0             & {\alpha} & {\beta}  \\
        0             & {\gamma} + {\delta} & {\gamma} & {\delta} \\
        {\alpha}  & {\gamma}  & {\alpha} + {\gamma} & 0      \\
        {\beta}   & {\delta}  & 0               & {\beta} + {\delta}
    \end{bmatrix}.
    \label{eq:fisher_4eq}
\end{equation}
The closed-form estimates of Alice's and Trudy's fingerprints are the
second and third entries of $\hat{\boldsymbol{\theta}}_k$ respectively:
\begin{align}
    \hat{{f}}_{A,k} &= \left[\hat{\boldsymbol{\theta}}_k\right]_2
    = \left[{\bm{J}}^{-1} \bm{H}^H {\boldsymbol{\Sigma}}^{-1}
    \bar{\bm{f}}_k\right]_2, \label{eq:falice_estimate} \\
    \hat{{f}}_{T,k} &= \left[\hat{\boldsymbol{\theta}}_k\right]_3
    = \left[{\bm{J}}^{-1} \bm{H}^H {\boldsymbol{\Sigma}}^{-1}
    \bar{\bm{f}}_k\right]_3. \label{eq:ftrudy_estimate}
\end{align}

\subsubsection*{Estimation Variance}

The estimation error $\boldsymbol{\epsilon} = \hat{\boldsymbol{\theta}}_k
- \boldsymbol{\theta}_k$ is zero-mean complex Gaussian with covariance:
\begin{equation}
    \bm{C}_{\boldsymbol{\theta}} = {\bm{J}}^{-1} =
    \left(\bm{H}^H {\boldsymbol{\Sigma}}^{-1} \bm{H}\right)^{-1}.
\end{equation}
The variances of the two estimates of interest are the $(2,2)$ and $(3,3)$ diagonal entries of $\bm{C}_{\boldsymbol{\theta}}$, obtained as:
\begin{align}
    \sigma^2_{A,k} &= \left[\bm{C}_{\boldsymbol{\theta}}\right]_{2,2}
    = \frac{C_{22}}{\det({\bm{J}})}, \label{eq:var_alice} \\
    \sigma^2_{T,k} &= \left[\bm{C}_{\boldsymbol{\theta}}\right]_{3,3}
    = \frac{C_{33}}{\det({\bm{J}})}, \label{eq:var_trudy}
\end{align}
where $C_{22}$ and $C_{33}$ are the $(2,2)$ and $(3,3)$ cofactors of
 ${\bm{J}}$, i.e.\ the determinants of the $3\times 3$ submatrices
obtained by deleting row 2, column 2 and row 3, column 3 from
 ${\bm{J}}$ respectively. Substituting the entries
of~\eqref{eq:fisher_4eq}:
\begin{equation}
    \det({\bm{J}}) = ({\alpha}{\delta} - {\beta}{\gamma})^2,
\end{equation}
\begin{multline}
    C_{22} = ({\alpha} + {\gamma})({\beta} + {\delta})({\alpha} + {\beta})
    - {\alpha}^2({\beta} + {\delta}) \\
    - {\beta}^2({\alpha} + {\gamma}),
\end{multline}
\begin{multline}
    C_{33} = ({\alpha} + {\beta})({\gamma} + {\delta})({\beta} + {\delta})
    - {\beta}^2({\gamma} + {\delta}) \\
    - {\delta}^2({\alpha} + {\beta}).
\end{multline}

\paragraph*{Remark} We here recall that the presence of Chuck is a strict
requirement for the attack: since Trudy transmits the spoofing signal
 $\bm{s}$, her own hardware fingerprint ${f}_{T,k}$ is inevitably
imprinted on the transmission and must be estimated and then pre-canceled
while constructing $\bm{s}$. With this attack framework, we then use
the minimum number required to reach a successful attack as we give to
Trudy one colluding ally.

\subsection{Trudy's Forged Signal}
Bob receives Trudy's spoofing signal through the true channel $\bm{h}_{TB}$ and Trudy's
true hardware fingerprint ${\bm{f}}_T$. The \ac{csi} at Bob on subcarrier $k$ is \eqref{eq:csi_bob_raw}.
After Bob applies his estimation and divides by 
his channel estimate $\hat{h}_{TB,k} = h_{TB,k} + \varepsilon_{est,k}$, 
the extracted fingerprint on subcarrier $k$ becomes:
\begin{multline}
    \hat{f}_{TB,k} = (1 + {f}_{T,k} + f_{B,k})  s_{k} + 
    \frac{z_{B,k}}{h_{TB,k}} \\
    = \frac{(1 + {f}_{A,k} + \varepsilon_{A,k})(1 + {f}_{T,k} + f_{B,k})}
    {(1 + \hat{f}_{T,k})  \hat{h}_{TB,k}}
    + \frac{z_{B,k}}{h_{TB,k}}.
    \label{eq:bob_extracted}
\end{multline}
We now use the definitions $\hat{{f}}_{T,k} = 1 + {f}_{T,k} + \varepsilon_{T,k}$,
 $\hat{h}_{TB,k} = h_{TB,k} + \varepsilon_{\text{est},k}$, and
 $\hat{{f}}_{A,k} = 1 + {f}_{A,k} + \varepsilon_{A,k}$ in \eqref{eq:bob_extracted}, where
 $\varepsilon_{A,k} \sim \mathcal{CN}(0, \sigma^2_{A,k})$ is the \ac{wls}
estimation error from~\eqref{eq:var_alice}. With these definitions we can rewrite \eqref{eq:bob_extracted} as
\begin{equation}
    \hat{f}_{TB,k} =
    \frac{(1 + {f}_{A,k} + \varepsilon_{A,k})
     (1 + {f}_{T,k} + f_{B,k})}
    {\left[(1 + {f}_{T,k}) + \varepsilon_{T,k}\right]
     \left[h_{TB,k} + \varepsilon_{\text{est},k}\right]}.
    \label{eq:bob_extracted_full}
\end{equation}
Factoring $(1 + f_{T,k})$ and $h_{TB,k}$ out of the denominator and applying a first-order Taylor expansion, valid when 
 $|\varepsilon_{T,k}| \ll |1 + {f}_{T,k}|$ and 
 $|\varepsilon_{est,k}| \ll |h_{TB,k}|$, and retaining only 
first-order terms:
\begin{align}
    \hat{f}_{TB,k} \approx\; & (1 + {f}_{A,k} + f_{B,k}) \notag\\
    &- \underbrace{\frac{1 + {f}_{A,k} + f_{B,k}}{1 + {f}_{T,k}}
    \varepsilon_{T,k}}_{\Delta^{T}_k}
    - \underbrace{\frac{1 + {f}_{A,k} + f_{B,k}}{h_{TB,k}}
    \varepsilon_{\text{est},k}}_{\Delta^{\text{est}}_k} \notag\\
    &- \underbrace{\frac{z_{B,k}}{h_{TB,k}}}_{\Delta^{\text{Bob}}_k}
    + \underbrace{\varepsilon_{A,k}}_{\Delta^{A}_k}.
    \label{eq:bob_distortion}
\end{align}
A detailed derivation of \eqref{eq:bob_distortion} is presented in Appendix~\ref{app::3}.
The total distortion $\boldsymbol{\Delta} = \boldsymbol{\Delta}^{T} +
\boldsymbol{\Delta}^{\text{est}} + \boldsymbol{\Delta}^{\text{Bob}} +
\boldsymbol{\Delta}^{A}$ is the sum of four independent zero-mean complex
Gaussian vectors, since $\varepsilon_{T,k}$, $\varepsilon_{\text{est},k}$,
 $z_{B,k}$, and $\varepsilon_{A,k}$ are all independent. The variances of each 
contribution are:
\begin{equation}
    \text{Var}[\Delta^{T}_k] = \frac{|1 + f_{A,k} + f_{B,k}|^2}{|1+f_{T,k}|^2}\sigma^2_{T,k},
\end{equation}
\begin{equation}
    \text{Var}[\Delta^{\text{est}}_k] = \frac{|1 + f_{A,k} + f_{B,k}|^2}{|h_{TB,k}|^2}\sigma^2_{\text{est},k},
\end{equation}
\begin{equation}
    \text{Var}[\Delta^{\text{Bob}}_k] = \frac{\sigma^2_B}{|h_{TB,k}|^2}, \qquad
    \text{Var}[\Delta^{A}_k] = \sigma^2_{A,k}.
\end{equation}
Note that unlike $\Delta^{T}_k$ and $\Delta^{\text{est}}_k$, the term $\Delta^{A}_k =
\varepsilon_{A,k}$ does not carry the $|1 + f_{A,k} + f_{B,k}|^2$ scaling
factor, since the \ac{wls} estimation error enters additively in the numerator
of~\eqref{eq:spoofing_signal} rather than multiplicatively through the
fingerprint pre-cancellation.

\section{Security Analysis}
\label{sec::test_function}
In this section, we analytically derive the performance of the \ac{mcsi}-based authentication system under the proposed colluding attack. In particular, we derive an approximation of the distributions of the authentication test statistic under both the legitimate and attack hypotheses, enabling the derivation of a closed-form formula of the false alarm probability and a numerical computation of the misdetection probability.

\subsection{False Alarm Probability}

Under $\mathcal{H}_0$, Bob receives a legitimate transmission from Alice.
The extracted fingerprint on subcarrier $k$ is compared against the stored
reference $f_{\text{Alice},k}$, which by definition is an estimate of the true
fingerprint ${f}_{A,k}$ affected by noise during the enrollment process, thus different.
The difference after the estimation process is then:
\begin{multline}
\hat{f}_{TB,k} - f_{\text{Alice},k} =
\underbrace{\left(1 + {f}_{A,k} + {f}_{B,k} +  \varepsilon_k\right)}_{\text{live estimation}} \\
- \underbrace{\left(1 + {f}_{A,k} + {f}_{B,k} + \varepsilon_{\text{Bob},k}\right)}_{\text{stored reference}} \\
= \varepsilon_k - \varepsilon_{\text{Bob},k},
\end{multline}
where $\varepsilon_k \sim \mathcal{CN}(0, \sigma^2_k)$ is the residual estimation
noise on subcarrier $k$ after estimation, and $\varepsilon_{\text{Bob},k} \sim
\mathcal{CN}(0, \sigma^2_{\text{E},k})$ is the enrollment noise with:
\begin{equation}
    \sigma^2_{\text{E},k} = \frac{\sigma^2_B}{N_{\text{E}}\,|h_{AB,k}|^2}.
\end{equation}
Since $\varepsilon_k$ and $\varepsilon_{\text{Bob},k}$ are independent, their difference
is zero-mean complex Gaussian with combined variance:
\begin{equation}
    \nu^2_k \triangleq \sigma^2_k + \sigma^2_{\text{E},k}.
    \label{eq:nu_k}
\end{equation}

Note that when the live estimation and enrollment share the same channel 
 $h_{AB,k}$ and noise variance $\sigma_B^2$, the two contributions in 
\eqref{eq:nu_k} simplify to:
\begin{align}
    \nu_k^2 &= \frac{\sigma_B^2}{N_{\text{A}}|h_{AB,k}|^2} + 
    \frac{\sigma_B^2}{N_{\text{E}}|h_{AB,k}|^2} \nonumber \\
    &= \frac{\sigma_B^2}{|h_{AB,k}|^2}
    \left(\frac{1}{N_{\text{A}}} + \frac{1}{N_{\text{E}}}\right). \label{eq:nu_simplified}
\end{align}

By normalizing each subcarrier separately in the test statistic \eqref{eq:glrt}, the real and imaginary parts act as independent standard normal random variables.
The per-subcarrier contribution of the sum in \eqref{eq:glrt} is
\begin{multline}
    \frac{2|\varepsilon_k - \varepsilon_{\text{Bob},k}|^2}{\nu_k^2}
    = \left(\frac{\mathrm{Re}(\varepsilon_k - \varepsilon_{\text{Bob},k})}
    {\sqrt{\nu_k^2/2}}\right)^2 \\
    + \left(\frac{\mathrm{Im}(\varepsilon_k - \varepsilon_{\text{Bob},k})}
    {\sqrt{\nu_k^2/2}}\right)^2
    \approx \chi^2_2,
\end{multline}
where each squared term is the square of a standard normal by construction and $\chi_n^2$ is the chi-square distribution of order $n$.
By summing over all $K$ subcarriers, we have:
\begin{equation}
    \Psi\big|_{\mathcal{H}_0} = 2\sum_{k=1}^{K}
    \frac{|\varepsilon_k - \varepsilon_{\text{Bob},k}|^2}{\nu^2_k}
    \approx \chi^2_{2K}.
    \label{eq:h0_stat}
\end{equation}
Crucially, this result
holds for any values of $\sigma^2_k$ and $\sigma^2_{\text{E},k}$ across subcarriers: by doing 
per-subcarrier normalization $\nu^2_k$ we end up in a distribution which is a standard central $\chi^2$ with $2K$ degrees of freedom regardless.
The threshold $\tau$ and the false alarm probability are
therefore related analytically by:
\begin{equation}
    P_{FA}(\tau) = P\!\left(\Psi\big|_{\mathcal{H}_0} > \tau\right)
    = 1 - F_{\chi^2_{2K}}(\tau),
    \label{eq:pfa}
\end{equation}
where $F_{\chi^2_{2K}}()$ is the \ac{cdf} of the central $\chi^2$
distribution with $2K$ degrees of freedom, and $\tau$ can be set
 to meet any target false alarm probability through the inversion of \eqref{eq:pfa}.

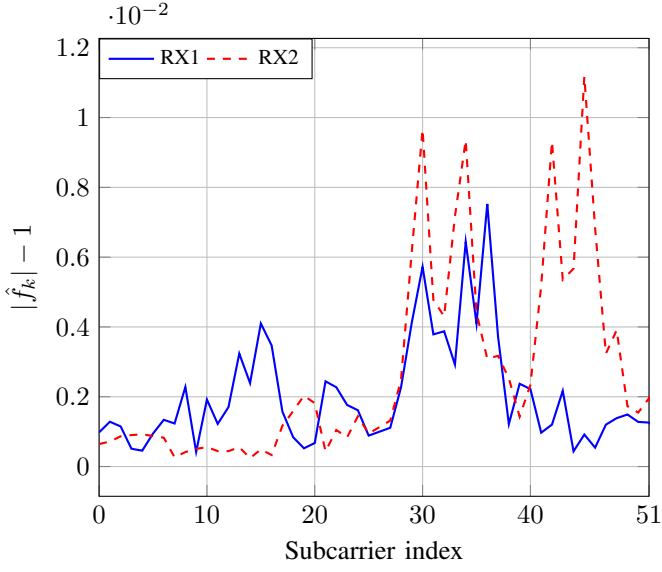
\begin{figure}
\centering
\begin{tikzpicture}
    \begin{axis}[
        width=\columnwidth,
        xlabel={Subcarrier index},
        ylabel={$|\hat{f}_k|-1$}, 
        grid=both,
        xmin=0, xmax=51,
        xtick={0, 10, 20, 30, 40, 51},
        legend style={font=\footnotesize, at={(0,1)}, anchor=north west, legend columns=2},
    ]
 
    \addplot[color=blue, solid, thick] table [col sep=comma, x expr=\thisrow{subcarrier}+26, y expr=\thisrow{magnitude_rx1}] {Data/fingerprints_magnitude.csv};
    \addlegendentry{RX1}
    \addplot[color=red, dashed, thick] table [col sep=comma, x expr=\thisrow{subcarrier}+26, y expr=\thisrow{magnitude_rx2}] {Data/fingerprints_magnitude.csv};
    \addlegendentry{RX2}
 
    \end{axis}
\end{tikzpicture}
\caption{Magnitude of the received composite fingerprint for the same transmitter at two different receivers.}
\label{fig:magnitude}
\end{figure}
\subsection{Misdetection Probability}
Under $\mathcal{H}_1$, Bob receives Trudy's spoofed transmission.
From~\eqref{eq:bob_distortion}, the residual on subcarrier $k$ is
 $\Delta_k \sim \mathcal{CN}(0, \mathrm{Var}[\Delta_k])$, so the
per-subcarrier contribution to the test statistic~\eqref{eq:glrt} is:
\begin{equation}
    \frac{2|\Delta_k|^2}{\nu^2_k} =
    \underbrace{\frac{\mathrm{Var}[\Delta_k]}{\nu^2_k}}_{c_k}
     \underbrace{\frac{2|\Delta_k|^2}{\mathrm{Var}[\Delta_k]}}_{\approx\,\chi^2_2}
    = c_k  \chi^2_2.
    \label{eq:h1_per_sub}
\end{equation}
Defining
\begin{align}
c_k &= \frac{\mathrm{Var}[\Delta_k]}{\nu_k^2} \notag\\
&= \frac{1}{\nu_k^2}\Bigg(
\frac{|1+{f}_{A,k}+f_{B,k}|^2\,\sigma^2_{T,k}}{|1+{f}_{T,k}|^2}
+ \frac{|1+{f}_{A,k}+f_{B,k}|^2\,\sigma^2_{\text{est},k}}{|h_{TB,k}|^2} \notag\\
&\quad + \frac{\sigma^2_B}{|h_{TB,k}|^2}
+ \sigma^2_{A,k}
\Bigg),
\label{eq:weights_new}
\end{align}
the test statistic under $\mathcal{H}_1$ can be written as 
\begin{equation}
    \Psi\big|_{\mathcal{H}_1} = 2\sum_{k=1}^{K}
    \frac{|\Delta_k|^2}{\nu^2_k} = \sum_{k=1}^{K}
    c_k u_k \sim \bar{\chi}^2_{2K}(\bm{c}_k),
    \label{eq:psi_h1_weighted}
\end{equation}
where $u_k \sim \mathcal{CN}(0, 1)$, and $\bar{\chi}^2_n(\bm{x})$ is a generalized $\chi^2$ distribution with $n$ degrees of freedom and weights in vector $\bm{x}$. Note that there is no closed-form expression for the \ac{cdf} of a generalized $\chi^2$ distribution, but it can be computed via Imhof's numerical method \cite{imhof_book}.

\begin{figure}
\centering
\begin{tikzpicture}
    \begin{axis}[
        width=\columnwidth,
        xlabel={Subcarrier index},
        ylabel={$\angle\hat{f}_k$ (deg)},
        ymin=-180, ymax=180,
        ytick={-180,-90,0,90,180},
        xmin=0, xmax=51,
        xtick={0, 10, 20, 30, 40, 51},
        grid=both,
        legend style={font=\footnotesize, at={(0,1)}, anchor=north west, legend columns=2},
    ]

    \addplot[color=blue, solid, thick] table [col sep=comma, x expr=\thisrow{subcarrier}+26, y=phase_deg_rx1] {Data/fingerprints_phase.csv};
    \addlegendentry{RX1}

    \addplot[color=red, dashed, thick] table [col sep=comma, x expr=\thisrow{subcarrier}+26, y=phase_deg_rx2] {Data/fingerprints_phase.csv};
    \addlegendentry{RX2}

    \end{axis}
\end{tikzpicture}
\caption{Phase of the received \ac{csi} for the same transmitter at two different receivers.}
\label{fig:phase}
\end{figure}
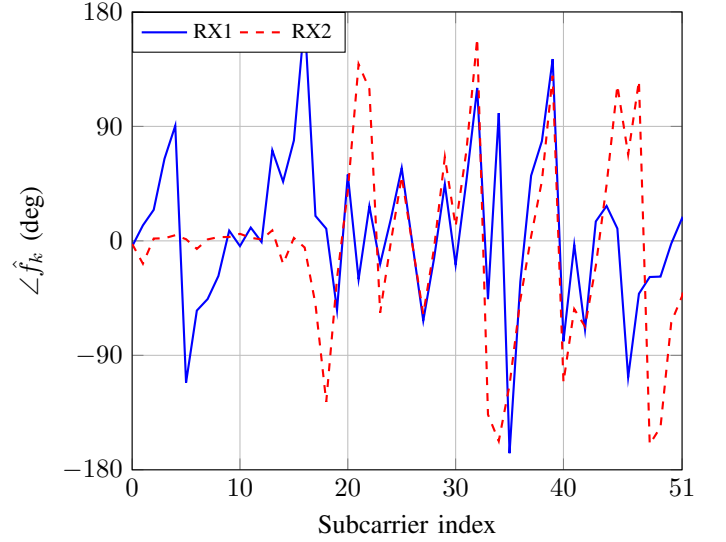

\section{Numerical Results}
\label{sec::numerical}
We evaluate the performance of the proposed authentication system and the effectiveness of the spoofing attack using the WiSig RF fingerprinting dataset \cite{wisig}, which includes $174$ Wi-Fi devices as transmitters and $41$ USRP devices as receivers. In the dataset, the CSI is computed using $52$ pilots with the long training field sequence over a $20$\, MHz of bandwidth. Note that in this case we set $N_p=8$ to be consistent with \cite{chen}.

The analytical distributions derived in Section~\ref{sec::test_function}, represented with solid lines, are validated against Monte-Carlo simulations of the attack using the \ac{mcsi} fingerprints extracted from the dataset.

\subsection{Receiver-Dependent \ac{mcsi} Characteristics}

Before analyzing authentication performance, we first validate a fundamental assumption of our model, so that the extracted \ac{mcsi} fingerprint conflates both transmitter and receiver hardware impairments. Fig.s~\ref{fig:magnitude} and~\ref{fig:phase} illustrate the magnitude and phase of the estimated \ac{mcsi} $\hat{f}_k$ for the same transmitting device observed at two different receivers. Note that both magnitude and phase exhibit different across all subcarriers, proving the receiver-specific distortions. This confirms that the composite fingerprint $\bm{f}$ is indeed receiver-dependent, justifying our modeling choice to treat ${f}^{\text{RX}}$ as a constant that cancels during legitimate authentication but must be explicitly estimated by an attacker attempting spoofing.

\subsection{Area Under The Curve Performance}

We now show the security performance of \ac{mcsi}-based authentication. We resort to the \ac{auc}, which is defined as the integral of the detection error tradeoff (DET) curve, plotting the misdetection probability ($P_{MD}$) against the false alarm probability ($P_{FA}$) as the decision threshold $\tau$ varies. The \ac{auc} provides a standard security metric of the test's performance, ranging from $0$ (perfect detection - best case for the defense) to $0.5$ (random guessing - best case for the attacker).

Fig.~\ref{fig::auc_1} shows the \ac{auc} as a function of different values of $M_i$, for different values of the enrollment count $N_{\rm E}$ and fixed $\sigma^2_{\rm B}=10^{-3}$, $N_{\rm A}=1$ and all of the noise variances $\sigma^2_{BT}=\sigma^2_{BC}=\sigma^2_{AT}=\sigma^2_{AC}=1.0$. 
By being all equally set in the simulations, we will refer to the noise variance at Trudy's receiver as $\sigma^2_{\rm{T}}$ and to the noise variance at Chuck's receiver as $\sigma^2_{\rm{C}}$.
As the attacker weight increases, corresponding to Trudy accumulating more pilot observations and thus reducing its estimation variance, the \ac{auc} monotonically tends toward $0.5$, which corresponds to a random-guess classifier unable to distinguish legitimate from spoofed transmissions, and so the worst-case scenario for the defense. This confirms that, given sufficient observations, even if the attacker has a noise variance that can be orders of magnitude higher compared to the one of Bob, Trudy can forge Alice's individual fingerprint with enough accuracy to render Bob's authentication test statistically indistinguishable from the legitimate case. 

The enrollment count $N_{\rm E}$ shifts the \ac{auc} curve, as a larger stored reference reduces Bob's enrollment noise $\sigma^2_{\rm E}$ and tightens the authentication threshold; however, this advantage is overcome as the attacker weight grows, and the \ac{auc} ultimately converges to $0.5$ regardless of $N_{\rm E}$.
Fig.~\ref{fig::auc_2} displays this behavior from the perspective of Bob's noise $\sigma^2_{\rm B}$, with fixed $N_{\rm E}=20$, $N_{\rm A}=1$, and the same values of noise variances at Trudy's and Chuck's receivers. As we can see, larger $\sigma^2_{\rm B}$ degrades the quality of the live fingerprint estimation and loosens Bob's decision boundary, making the system easier to fool even at lower attacker weights. Conversely, a lower $\sigma^2_{\rm B}$ delays the convergence of \ac{auc} to $0.5$ but does not avoid it: for any fixed noise level at Bob, a sufficiently capable attacker will always make the \ac{auc} converge to $0.5$.

At the end, we note that in Fig. \ref{fig::auc_1} and Fig. \ref{fig::auc_2} the solid curves computed using the theoretical derivations found in Sec.~\ref{sec::test_function} almost perfectly fit the Monte-Carlo simulations of the attack, showing a correspondence between the theoretical derivation and the attack simulations.

\subsection{Detection Error Tradeoff Performance}
For completeness, we also display individual \ac{det} curves for a changing parameter computed using Monte-Carlo simulations.

Fig.~\ref{fig::det_1} shows the \ac{det} curves for different values of $\sigma^2_{\rm T}$, with fixed $\sigma^2_{\rm B}=0.1$, $\sigma^2_{\rm C}=0.1$, $N_{\rm A}=1$, and $N_{\rm E}=20$. As $\sigma^2_{\rm T}$ decreases, the \ac{det} curve approaches the diagonal of the unit square, indicating that no operating point of the threshold $\tau$ allows the authenticator to simultaneously achieve low $P_{\rm FA}$ and low $P_{\rm MD}$. In particular, when $\sigma^2_{\rm T}$ matches the noise level at Bob, the \ac{det} curve lies on the diagonal, meaning that Bob becomes a random classifier. This result directly implies that, under equal noise conditions, \ac{mcsi}-based authentication provides no meaningful security guarantee against a smart adversary equipped with the colluder-assisted spoofing strategy of Section~\ref{sec::attacker}.

\begin{figure}
\centering
\begin{tikzpicture}
    \begin{axis}[
        width=\columnwidth,
        xlabel={$P_{\mathrm{FA}}$},
        ylabel={$P_{\mathrm{MD}}$},
        xmin=0, xmax=1,
        ymin=0, ymax=1,
        grid=both,
        legend style={font=\footnotesize, at={(1,1)}, anchor=north east, legend columns=2},
    ]

    \addplot+ [smooth, mark=none, line width=1pt] table [col sep=comma, x=P_FA, y={PMD_Navg1}] {Data/DET_Navg_sweep.csv};
    \addlegendentry{$1$}

    \addplot+ [smooth, mark=none, line width=1pt] table [col sep=comma, x=P_FA, y={PMD_Navg5}] {Data/DET_Navg_sweep.csv};
    \addlegendentry{$5$}

    \addplot+ [smooth, mark=none, line width=1pt] table [col sep=comma, x=P_FA, y={PMD_Navg10}] {Data/DET_Navg_sweep.csv};
    \addlegendentry{$10$}

    \addplot+ [smooth, mark=none, line width=1pt] table [col sep=comma, x=P_FA, y={PMD_Navg20}] {Data/DET_Navg_sweep.csv};
    \addlegendentry{$20$}

    \end{axis}
\end{tikzpicture}
\caption{\ac{det} curve for different values of $N_{\rm A} = 1, 5, 10$, and 20, with fixed $\sigma^2_{\rm{B}}=\sigma^2_{\rm{T}}=\sigma^2_{\rm{C}}=0.1$ and $N_{\rm{E}}=20$.}
\label{fig::det_2}
\end{figure}
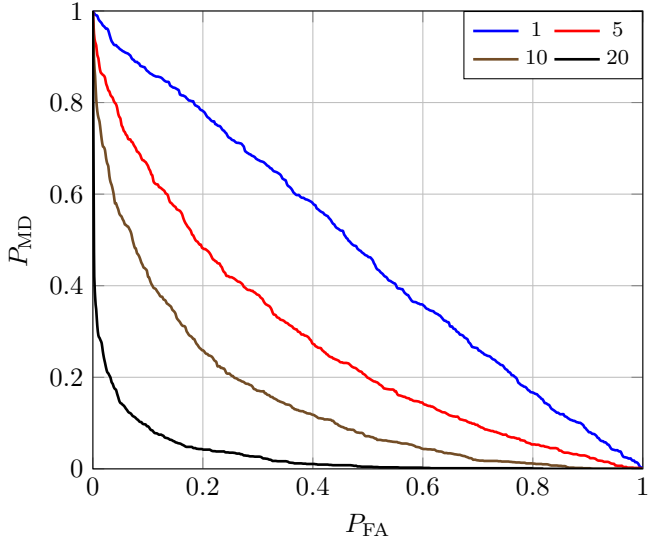

Fig.~\ref{fig::det_2} further shows that increasing $N_{\rm A}$ improves the defender's operating point for a fixed noise budget ($\sigma^2_{\rm B}=\sigma^2_{\rm T}=\sigma^2_{\rm T}=0.1$, $N_{\rm E}=20$), shifting the \ac{det} curve away from the diagonal. However, this gain does not come for free, as we demonstrate in this work that a smarter attacker can equally exploit this by averaging multiple observations on its side, as shown in Fig. \ref{fig::auc_1} and \ref{fig::auc_2}. The fundamental conclusion is therefore that \ac{mcsi}, while a simple and stable feature for closed-set device identification, does not constitute a secure basis for authentication in the presence of an intelligent adversary, as the security guarantees collapse when an attacker is granted sufficient observations of the legitimate channel or is affected by the same noise levels as the legitimate devices.

\begin{figure}
\centering
\begin{tikzpicture}
    \begin{axis}[
        xmode=log,
        width=\columnwidth,
        xlabel={$M_i$},
        ylabel={AUC},
        grid=both,
        legend style={font=\footnotesize,at={(0,1)},anchor=north west,legend columns=2},
        cycle list={{blue,mark=*},{red,mark=*},{green!60!black,mark=*},{orange,mark=*}},
    ]
    
    \addplot+[no markers, solid, color=blue] 
        table [col sep=comma, x=x, y=auc_Ne_1] {Data/imhof_auc_data.csv};
    \addlegendentry{$1$}
    \addplot+[only marks, mark=*, color=blue, forget plot] 
        table [col sep=comma, x=x_val, y=auc_Ne_1] {Data/auc_data.csv};
    
    \addplot+[no markers, solid, color=red] 
        table [col sep=comma, x=x, y=auc_Ne_5] {Data/imhof_auc_data.csv};
    \addlegendentry{$5$}
    \addplot+[only marks, mark=*, color=red, forget plot] 
        table [col sep=comma, x=x_val, y=auc_Ne_5] {Data/auc_data.csv};
    
    \addplot+[no markers, solid, color=green!60!black] 
        table [col sep=comma, x=x, y=auc_Ne_20] {Data/imhof_auc_data.csv};
    \addlegendentry{$20$}
    \addplot+[only marks, mark=*, color=green!60!black, forget plot] 
        table [col sep=comma, x=x_val, y=auc_Ne_20] {Data/auc_data.csv};
    
    \addplot+[no markers, solid, color=orange] 
        table [col sep=comma, x=x, y=auc_Ne_50] {Data/imhof_auc_data.csv};
    \addlegendentry{$50$}
    \addplot+[only marks, mark=*, color=orange, forget plot] 
        table [col sep=comma, x=x_val, y=auc_Ne_50] {Data/auc_data.csv};
    
    \end{axis}
\end{tikzpicture}
\caption{\ac{auc} versus number of multiple measurements $M_i$ performed by Trudy and Chuck evaluated for different values of $N_{\text{E}} = 1, 5, 20,$ and $50$, for a fixed ($\sigma^2_{\rm{B}}=10^{-3}$). The dots represent the Monte-Carlo simulations result while the solid lines correspond to the theoretical curves.}
\label{fig::auc_1}
\end{figure}

\begin{figure}
\centering           
\begin{tikzpicture}
    \begin{axis}[
        xmode=log,
        width=\columnwidth, 
        xlabel={$M_i$},
        ylabel={AUC},
        grid=both,
        legend style={font=\footnotesize,at={(0,1)},anchor=north west,legend columns=2}, 
    ]

    \addplot[no markers, solid, color=green!60!black] 
        table [col sep=comma, x=x, y=auc_s2Bob_1.0e-02] {Data/imhof_auc_data_2.csv};
    \addlegendentry{$10^{-2}$}
    \addplot[only marks, mark=*, color=green!60!black, mark options={fill=green!60!black}, forget plot] 
        table [col sep=comma, x=x_val, y=auc_s2Bob_1.0e-02] {Data/auc_data_2.csv};
    
    \addplot[no markers, solid, color=orange] 
        table [col sep=comma, x=x, y=auc_s2Bob_1.0e-01] {Data/imhof_auc_data_2.csv};
    \addlegendentry{$10^{-1}$}
    \addplot[only marks, mark=*, color=orange, mark options={fill=orange}, forget plot] 
        table [col sep=comma, x=x_val, y=auc_s2Bob_1.0e-01] {Data/auc_data_2.csv};
    
    \addplot[no markers, solid, color=cyan] 
        table [col sep=comma, x=x, y=auc_s2Bob_1.0e+00] {Data/imhof_auc_data_2.csv};
    \addlegendentry{$1$}
    \addplot[only marks, mark=*, color=cyan, mark options={fill=cyan}, forget plot] 
        table [col sep=comma, x=x_val, y=auc_s2Bob_1.0e+00] {Data/auc_data_2.csv};
    
    \addplot[no markers, solid, color=magenta] 
        table [col sep=comma, x=x, y=auc_s2Bob_5.0e+00] {Data/imhof_auc_data_2.csv};
    \addlegendentry{$5$}
    \addplot[only marks, mark=*, color=magenta, mark options={fill=magenta}, forget plot] 
        table [col sep=comma, x=x_val, y=auc_s2Bob_5.0e+00] {Data/auc_data_2.csv};

    \end{axis}
\end{tikzpicture}
\caption{\ac{auc} versus number of multiple measurements $M_i$ performed by Trudy and Chuck evaluated for different values of $\sigma^2_{\rm{B}} = 0.01, 0.1, 1$, and $5$, for a fixed $N_{\rm E}=20$. The dots represent the Monte-Carlo simulation results while the solid lines correspond to the theoretical curves.}
\label{fig::auc_2}
\end{figure}
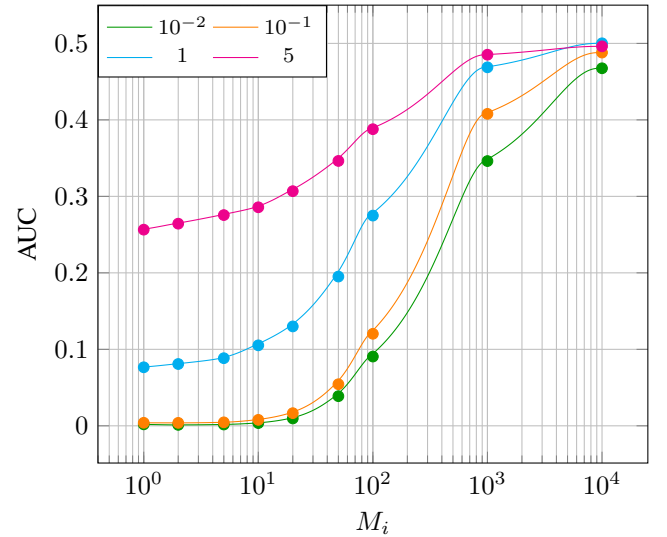

\begin{figure}
\centering
\begin{tikzpicture}
    \begin{axis}[
        width=\columnwidth,
        xlabel={$P_{\mathrm{FA}}$},
        ylabel={$P_{\mathrm{MD}}$},
        xmin=0, xmax=1,
        ymin=0, ymax=1,
        grid=both,
        legend style={font=\footnotesize, at={(1,1)}, anchor=north east, legend columns=2},
    ]
 
 
    \addplot+ [smooth, mark=none, line width=1pt] table [col sep=comma, x=P_FA, y={PMD_sT0.01}] {Data/DET_sigma_trudy_sweep.csv};
    \addlegendentry{$10^{-2}$}
 
    \addplot+ [smooth, mark=none, line width=1pt] table [col sep=comma, x=P_FA, y={PMD_sT0.1}] {Data/DET_sigma_trudy_sweep.csv};
    \addlegendentry{$10^{-1}$}
 
    \addplot+ [smooth, mark=none, line width=1pt] table [col sep=comma, x=P_FA, y={PMD_sT1}] {Data/DET_sigma_trudy_sweep.csv};
    \addlegendentry{$1$}
    
    \addplot+ [smooth, mark=none, line width=1pt] table [col sep=comma, x=P_FA, y={PMD_sT2}] {Data/DET_sigma_trudy_sweep.csv};
    \addlegendentry{$2$}
 
    \end{axis}
\end{tikzpicture}
\caption{\ac{det} curve for different values of $\sigma^2_{\rm{T}}= 0.01, 0.1, 1$, and 2, with fixed $\sigma^2_{\rm{B}}=\sigma^2_{\rm{C}}=0.1$ and $N_{\text{E}}=20$.}
\label{fig::det_1}
\end{figure}
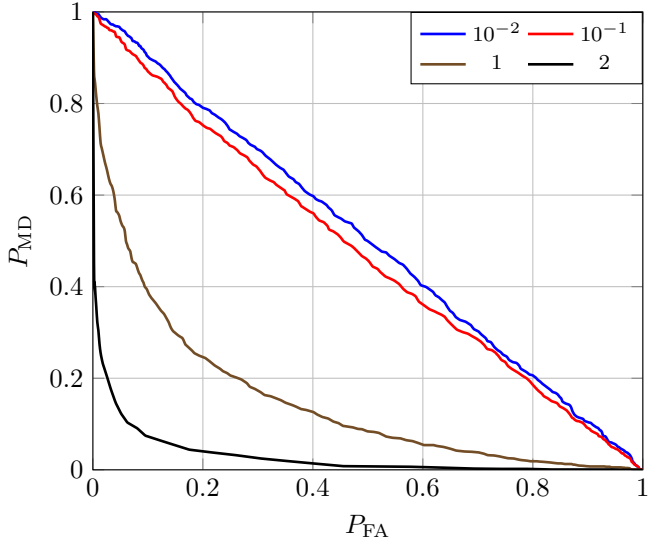

\section{Conclusions}
\label{sec::conclusion}

We have analytically characterized the security limits of \ac{mcsi}-based authentication in \ac{ofdm} systems against a smart adversary. By introducing a four-node scenario comprising Alice, Bob, Trudy, and the colluder Chuck, we designed an attack that jointly estimates the individual fingerprints of all devices via a \ac{wls} framework and constructs a forged signal able to fool authentication. The exact distributions of the authentication test statistic under both the legitimate and spoofing hypotheses were derived in closed form, enabling the analytical derivation of approximate false alarm and misdetection probabilities. The approximations have been confirmed to be tight by Monte-Carlo simulations.

The results reveal that \ac{mcsi}-based authentication provides no meaningful security guarantee against a sufficiently capable adversary: as the attacker accumulates pilot observations, the \ac{auc} monotonically converges to $0.5$ regardless of the enrollment count or the attacker's noise variance. In particular, under equal noise conditions between Bob and Trudy, the authentication system reduces to a random classifier.

\bibliographystyle{IEEEtran}
\bibliography{biblio.bib}

\appendices

\section{Derivation of \eqref{eq:f_alice}}
\label{app::1}

Bob collects $N_{\text{E}}$ independent transmissions from Alice on subcarrier $k$. On observation $i$, Bob receives the signal before pilot division:
\begin{equation}
    \begin{split}
        {r}^{(i)}_{B,k} &= h_{AB,k}(1 + {f}_{A,k}+{f}_{B,k}) x^{(i)}_k \\
        &\quad + \omega^{(i)}_{B,k},  \quad \omega^{(i)}_{B,k} \sim \mathcal{CN}(0, \sigma^2_\omega).
    \end{split}
\end{equation}
Dividing by the known pilot symbol $x^{(i)}_k$, the CSI on observation $i$ is obtained as:
\begin{equation}
    \begin{split}
        {c}^{(i)}_{B,k} &= \frac{{r}^{(i)}_{B,k}}{x^{(i)}_k} = h_{AB,k}(1 + {f}_{A,k}+{f}_{B,k}) \\
        &\quad + z^{(i)}_{B,k}, \quad z^{(i)}_{B,k} \sim \mathcal{CN}(0, \sigma^2_B).
    \end{split}
\end{equation}
Applying the estimation \eqref{eq::approximation}, Bob divides by the channel estimate $\hat{{h}}_k$, obtaining the composite fingerprint on observation $i$:
\begin{equation}
    \hat{{f}}^{(i)}_{AB,k} = \frac{{c}^{(i)}_{B,k}}{h_{AB,k}} \approx 1 + {f}_{A,k} + {f}_{B,k}+ \frac{z^{(i)}_{B,k}}{h_{AB,k}}.
\end{equation}
Bob then averages over all $N_{\text{E}}$ observations to form the stored reference:
\begin{multline}
    f_{\text{Alice},k} = \frac{1}{N_{\text{E}}} \sum_{i=1}^{N_{\text{E}}} \hat{{f}}^{(i)}_{AB,k} \\ 
    = 1 + {f}_{A,k} + {f}_{B,k}+ \frac{1}{N_{\text{E}}} \sum_{i=1}^{N_{\text{E}}} \frac{z^{(i)}_{B,k}}{h_{AB,k}}.
\end{multline}
Defining the enrollment noise term:
\begin{equation}
    \varepsilon_{Bob,k} = \frac{1}{N_{\text{E}}} \sum_{i=1}^{N_{\text{E}}} \frac{z^{(i)}_{B,k}}{h_{AB,k}},
\end{equation}
since the $z^{(i)}_{B,k}$ are i.i.d.\ $\mathcal{CN}(0, \sigma^2_B)$ and independent across observations, and $h_{AB,k}$ is a deterministic constant, $\varepsilon_{Bob,k}$ is zero-mean complex Gaussian with variance:
\begin{equation}
\begin{split}
    \text{Var}(\varepsilon_{Bob,k}) 
    & = \frac{1}{N^2_{\text{E}}} \sum_{i=1}^{N_{\text{E}}} \frac{\sigma^2_B}{|h_{AB,k}|^2} 
    = \frac{1}{N^2_{\text{E}}}  N_{\text{E}} \frac{\sigma^2_B}{|h_{AB,k}|^2} \\
    &     = \frac{\sigma^2_B}{N_{\text{E}} |h_{AB,k}|^2}.
\end{split}
\end{equation}
Therefore:
\begin{multline}
    f_{\text{Alice},k} = 1 + {f}_{A,k} + {f}_{B,k} + \varepsilon_{Bob,k}, \\
    \varepsilon_{Bob,k} \sim \mathcal{CN}\!\left(0,\, \frac{\sigma^2_B}{N_{\text{E}} |h_{AB,k}|^2}\right).
\end{multline}

\section{Derivation of \eqref{eq:trudy_channel_estimate}}
\label{app::2}

Trudy observes $L_T$ independent pilot transmissions from Bob on subcarrier $k$ as in \eqref{eq:trudy_observation}. Since $|f_{B,k}|, |f_{T,k}| \ll 1$ by the
approximation of Section~\ref{sec::system}, the multiplicative fingerprint
contribution satisfies $1 + f_{B,k} + f_{T,k} \approx 1$, and each observation
reduces to the fingerprint-free equivalent:
\begin{equation}
    r^{(l)}_{T,k} \approx h_{TB,k}  x^{(l)}_k + \omega^{(l)}_{T,k},
    \qquad l = 1, \ldots, L_T,
\end{equation}
where $x^{(l)}_k$ is the pilot symbol with power $P_B = \mathbb{E}[|x^{(l)}_k|^2]$ and $\omega^{(l)}_{T,k} \sim \mathcal{CN}(0, \sigma^2_T)$ are i.i.d. noise samples.
Dividing each observation by the known pilot symbol:
\begin{equation}
    c^{(l)}_{T,k} = \frac{r^{(l)}_{T,k}}{x^{(l)}_k} = h_{TB,k} + \frac{\omega^{(l)}_{T,k}}{x^{(l)}_k}
    \triangleq h_{TB,k} + z^{(l)}_{T,k},
\end{equation}
where $z^{(l)}_{T,k} = \omega^{(l)}_{T,k} / x^{(l)}_k$ is the normalized
noise term. Since $\omega^{(l)}_{T,k} \sim \mathcal{CN}(0, \sigma^2_T)$ and
 $|x^{(l)}_k|^2 = P_B$, the normalized noise has variance:
\begin{equation}
    \mathbb{E}\!\left[|z^{(l)}_{T,k}|^2\right]
    = \frac{\mathbb{E}[|\omega^{(l)}_{T,k}|^2]}{|x^{(l)}_k|^2}
    = \frac{\sigma^2_T}{P_B}.
\end{equation}
The LS estimator averages the $L_T$ normalized observations to minimize the
squared error:
\begin{equation}
    \hat{h}_{TB,k} = \frac{1}{L_T}\sum_{l=1}^{L_T} c^{(l)}_{T,k}
    = h_{TB,k} + \frac{1}{L_T}\sum_{l=1}^{L_T} z^{(l)}_{T,k}.
\end{equation}
The estimation error is therefore:
\begin{equation}
    \varepsilon_{\text{est},k} = \hat{h}_{TB,k} - h_{TB,k}
    = \frac{1}{L_T}\sum_{l=1}^{L_T} z^{(l)}_{T,k}.
\end{equation}
Since the $z^{(l)}_{T,k}$ are i.i.d. zero-mean complex Gaussian, the
error is also zero-mean complex Gaussian with variance:
\begin{align}
    \sigma^2_{\text{est},k}
    &= \mathbb{E}\!\left[|\varepsilon_{\text{est},k}|^2\right]
    = \frac{1}{L_T^2}\sum_{l=1}^{L_T}
      \mathbb{E}\!\left[|z^{(l)}_{T,k}|^2\right] \notag \\
    &= \frac{1}{L_T^2}  L_T  \frac{\sigma^2_T}{P_B}
    = \frac{\sigma^2_T}{L_T  P_B},
\end{align}
which ends the derivation of \eqref{eq:trudy_channel_estimate}.

\section{Derivation of \eqref{eq:bob_distortion}}
\label{app::3}
Before applying a first-order Taylor expansion to \eqref{eq:bob_extracted_full}, 
we first factor $(1+{f}_{T,k})$ and $h_{TB,k}$ out of the denominator:
\begin{multline}
    \hat{f}_{TB,k} = \frac{(1+{f}_{A,k} + \varepsilon_{A,k})  (1 + {f}_{T,k} + f_{B,k})}
    {(1+{f}_{T,k})  h_{TB,k}} \\
     \frac{1}{\left(1 + \frac{\varepsilon_{T,k}}{1+{f}_{T,k}}\right)
    \left(1 + \frac{\varepsilon_{\text{est},k}}{h_{TB,k}}\right)}.
\end{multline}
We define the shorthand $\phi_k \triangleq \frac{1+f_{T,k}+f_{B,k}}{1+f_{T,k}} \approx 1 + f_{B,k}$,
where the approximation follows from $|f_{T,k}| \ll 1$. The first fraction then simplifies to:
\begin{equation}
    \hat{f}_{TB,k} = \frac{(1+{f}_{A,k} + \varepsilon_{A,k})\,\phi_k}{h_{TB,k}}
     \frac{1}{\left(1 + \frac{\varepsilon_{T,k}}{1+{f}_{T,k}}\right)
    \left(1 + \frac{\varepsilon_{\text{est},k}}{h_{TB,k}}\right)}.
    \label{la}
\end{equation}
Under the conditions $|\varepsilon_{T,k}| \ll |1 + {f}_{T,k}|$ and 
 $|\varepsilon_{\text{est},k}| \ll |h_{TB,k}|$, we apply the first-order Taylor expansion 
 $1/(1+x) \approx 1-x$ to each factor and discard the second-order term that includes
 $\varepsilon_{T,k}  \varepsilon_{\text{est},k}$:
\begin{multline}\label{qui}
    \frac{1}{\left(1 + \frac{\varepsilon_{T,k}}{1+{f}_{T,k}}\right)
    \left(1 + \frac{\varepsilon_{\text{est},k}}{h_{TB,k}}\right)} \\
    \approx 1 - \frac{\varepsilon_{T,k}}{1+{f}_{T,k}} 
    - \frac{\varepsilon_{\text{est},k}}{h_{TB,k}}.
\end{multline}
Substituting \eqref{qui} in \eqref{la}, expanding, and retaining only first-order terms 
(i.e.\ dropping products of error terms such as 
 $\varepsilon_{A,k}  \varepsilon_{T,k}$, 
 $\varepsilon_{A,k}  \varepsilon_{\text{est},k}$, etc.) we obtain
\begin{align}
    \hat{f}_{TB,k} &\approx \frac{(1+{f}_{A,k} + \varepsilon_{A,k})\,\phi_k}{h_{TB,k}}
    \left(1 - \frac{\varepsilon_{T,k}}{1+{f}_{T,k}} 
    - \frac{\varepsilon_{\text{est},k}}{h_{TB,k}}\right) \notag\\
    &\approx \frac{(1+f_{A,k}+f_{B,k})}{h_{TB,k}}  h_{TB,k}
    - \frac{1+f_{A,k}+f_{B,k}}{1+{f}_{T,k}}\varepsilon_{T,k} \notag\\
    &\quad - \frac{1+f_{A,k}+f_{B,k}}{h_{TB,k}}\varepsilon_{\text{est},k}
    + \phi_k\,\varepsilon_{A,k}
\end{align}
where in the last step we used $\phi_k  h_{TB,k} / h_{TB,k} = \phi_k \approx 1+f_{B,k}$ and identified $(1+f_{A,k})(1+f_{B,k}) \approx 1+f_{A,k}+f_{B,k}$ since $|f_{A,k}|,|f_{B,k}| \ll 1$,
and similarly $\phi_k \approx 1$ for the $\varepsilon_{A,k}$ term to first order.
Adding the Bob receiver noise term $z_{B,k}/h_{TB,k}$ from \eqref{eq:bob_extracted}, 
we finally obtain
\begin{align}
    \hat{f}_{TB,k} \approx\; &(1+{f}_{A,k}+f_{B,k})  - \underbrace{\frac{1+{f}_{A,k}+f_{B,k}}{1 + {f}_{T,k}}
      \,\varepsilon_{T,k}}_{\Delta^{T}_k} \\ \notag
    & - \underbrace{\frac{1+{f}_{A,k}+f_{B,k}}{h_{TB,k}} \,\varepsilon_{\text{est},k}}_{\Delta^{\text{est}}_k} 
    - \underbrace{\frac{z_{B,k}}{h_{TB,k}}}_{\Delta^{\text{Bob}}_k}
    + \underbrace{\varepsilon_{A,k}}_{\Delta^{A}_k}.
    \label{eq:bob_distortion_appendix}
\end{align}

\end{document}